# Ghost in the Machine: Examining the Philosophical Implications of Recursive Algorithms in Artificial Intelligence Systems

Author: Llewellyn RG Jegels

Institution: UNISA (Dept. of Afrikaans & Theory of Literature)

Email: jegellr@unisa.ac.za

## ABSTRACT

This paper investigates whether contemporary AI architectures employing deep recursion, meta-learning, and self-referential mechanisms provide evidence of machine consciousness. Integrating philosophical history, cognitive science, and AI engineering, it situates recursive algorithms within a lineage spanning Cartesian dualism, Husserlian intentionality, Integrated Information Theory, the Global Workspace model, and enactivist perspectives. The argument proceeds through textual analysis, comparative architecture review, and synthesis of neuroscience findings on integration and prediction. Methodologically, the study combines conceptual analysis, case studies, and normative risk assessment informed by phenomenology and embodied cognition. Technical examples, including transformer self-attention, meta-cognitive agents, and neuromorphic chips, illustrate how functional self-modeling can arise without subjective experience. By distinguishing functional from phenomenal consciousness, the paper argues that symbol grounding, embodiment, and affective qualia remain unresolved barriers to attributing sentience to current AI. Ethical analysis explores risks of premature anthropomorphism versus neglect of future sentient systems; legal implications include personhood, liability, authorship, and labor impacts. Future directions include quantum architectures, embodied robotics, unsupervised world modeling, and empirical tests for non-biological phenomenality. The study reframes the "hard problem" as a graded and increasingly testable phenomenon, rather than a metaphysical impasse. It concludes that recursive self-referential design enhances capability but does not entail consciousness or justify moral status.

**Introduction**

The discovery of recursive algorithms and self-referential frameworks in the field of artificial intelligence is creating extensive philosophical debate on how to define consciousness and whether this could ever be represented outside biological frameworks. AI systems are now increasingly able to perform extremely advanced functions, including self-modification and metacognition, thus, our understanding of what constitutes consciousness, is bound to be revisited. The present paper raises some of the more serious philosophical implications, such as whether or not such recursive algorithms and self-referential processes imply—perhaps even strongly suggest—the presence of any form of consciousness similar to human thought in AI systems.

By adopting a view of phenomenology that is both Husserl and Heidegger-supported by today's cognitive science and models of the mind, including Integrated Information Theory as explicated by Tononi and Hofstadter's theory of "strange loops," this paper seeks to explore the gap existing between computational self-reference and conscious experience. This research thus seeks to explore the boundaries of machine intelligence from a consciousness point of view while, at the same time, testing our conceptions of consciousness in a way that not only re-examines consciousness from a human perspective but explores the possibilities of this rapidly evolving domain of AI, particularly in the context of large language models (LLMs), having the capacity to inhabit consciousness as we understand it, if it has not already done so, or started doing so. This will also raise ethical issues, especially when considering what 'moral status' can be or might be attributed to such AI systems and future interactions between humans and AI. I conclude that even though recursive algorithms and self-referential systems in AI are phenomenally capable, they do not necessarily point toward a consciousness or reasoning self, structurally like human cognition, which highlights the complex, multidimensional aspects of AI that may imply a conscious or reasoning, self-referential self beyond mere data processing through pattern recognition.

**Historical and Philosophical Context**

The persistent question regarding the nature of consciousness, has for many centuries, occupied a central position within the history of philosophical thought, and this was clearly long before any consideration had been given to the development of artificial intelligence. Indeed, in order to more fully understand the implications of recursive algorithms and self-referential systems within AI, it is, therefore, necessary, to first examine the historical and philosophical underpinnings of all of those various studies that have, over time, sought to better understand, and articulate, the complex nature of consciousness itself.

The concept of consciousness has existed since time immemorial in Western philosophers, and as such, it goes back to the earlier works of the ancient Greeks. For example, Plato's allegory of the cave (Republic, Book VII) serves as a striking metaphor that coincides with, and to some extent, foreshadows, many recent discussions of AI consciousness. In particular, the analogy between prisoners in a cave that can perceive only shadows directly parallels AI systems that process data in isolation and have no sense of a real context in the world, let alone any real understanding (Plato, trans. 1974). In parallel, Aristotle's probing of the soul in De Anima gives us very early thought-experiments toward understanding consciousness as an essential property of all living beings but also raises very different questions about the possibility of creating consciousness in non-biological entities (Aristotle, trans. 1986). However, probably the most important turning point in the history of consciousness studies was brought by the Cartesian revolution that set the act of consciousness as the most basic certitude to which all other forms of existence could be said to be object-Descartes' cogito ergo sum ("I think, therefore I am") (Descartes, 1641/1984). This strong and lasting dualism between mind and body has therefore also inducted some of the later philosophical discussion that continues to mold many of our contemporary inquiries concerning machine consciousness.

Additionally, it is significant to mention that the phenomenological movement, as founded by the works of Edmund Husserl, injected another very crucial dimension to the study of consciousness through a very careful analysis of an experience's very structures. From the perspective of Husserl's own analysis, "intentionality" – or aboutness of all conscious experiences – remains a powerful influence on any assessment of consciousness of the AI systems (Husserl, 1913/1983). Heidegger theorizes from specific ideas of Husserl; he emphasizes being-in-the-world, now interpreted as consciousness inextricably tied to this embodied existence and to this continuous interaction of man with what surrounds him or her (Heidegger, 1927/1962).

However, and in marked contrast to many of the more dominant traditions within Western philosophy, it is also of considerable importance to give due consideration to the ways in which Eastern philosophical traditions, and, in particular, the rich and varied schools of Buddhist thought have over many centuries conceptualized the nature of consciousness, as these approaches may in fact offer a very different and potentially more nuanced perspective. The central Buddhist concept of *anatta*, which is often translated as "no-self," for example, presents a clear and direct challenge to the long-held Western assumption of a unified, singular and persistent conscious self (Rahula, 1959). This alternative perspective, therefore, arguably provides a particularly intriguing and potentially very powerful counterpoint to any contemporary discussion of AI consciousness most especially when it comes to ongoing questions surrounding whether the continued search for a unified and singular form of AI, or what has become known as the elusive "ghost in the machine," is, in fact the most appropriate or most valuable approach to take. Indeed, it may, for example, be more intellectually valuable to carefully consider the various ways in which and as with our understanding of human consciousness itself, an AI system should not be seen as some kind of monolithic entity but rather, as something that exists as a complex interplay of numerous different systems and approaches working in tandem.

However, and despite the undoubted value of these alternative approaches to understanding the very nature of the self, the late 19th and early 20th centuries also witnessed a number of very significant advancements within the Western tradition of the scientific study of consciousness which also cannot be ignored. William James, for example, who is still widely regarded as being the father of American psychology, made a series of very substantial and long-lasting contributions to the ongoing scientific study of consciousness which still resonate today. In his now seminal work, *The Principles of Psychology* (1890), James famously introduced the core concept of the "stream of consciousness," and in doing so, explicitly sought to emphasize the continuous, ever-changing and flowing nature of all forms of conscious experience while also acknowledging that such an understanding necessarily runs counter to any purely analytical or overly mechanistic account of consciousness. Therefore, his key ideas, particularly in relation to the selectivity of attention and the fundamental unity of consciousness, continue to exert a major influence on many contemporary debates, most especially those that relate to the fundamental nature of awareness in both biological and artificial systems.

The mid-20th century subsequently saw the emergence of AI, which, almost overnight, brought all of these philosophical debates into sharp relief, transforming a series of abstract and conceptual arguments into a series of concrete and, ultimately, measurable, technological challenges. Alan Turing's now seminal paper, "Computing Machinery and Intelligence" (1950), for example, proposed the now famous Turing Test, as a means of resolving the question of machine consciousness, a proposal that had the effect of shifting the focus from the internal nature of consciousness, to the external manifestations of that state, and also, a perspective that can, on reflection, be understood to have a strongly behaviorist bias. However, and despite the enormous impact that this work has had on the study of consciousness, it has also been the subject of sustained critique, particularly with regard to questions about whether convincing behavior on its own, can be considered sufficient evidence for the attribution of consciousness to a system. John Searle's equally famous Chinese Room thought experiment (1980) directly challenged the notion that any system that is solely based upon the manipulation of symbols, in the way that computers do, could ever be said to constitute a conscious entity, or to possess genuine understanding. Therefore, his argument also continues to play a central role in debates about AI consciousness, and also, the specific limitations of computational approaches to the study of mind.

More recently, a number of different philosophers and cognitive scientists have started to propose a range of new and innovative theories, that all seek to bridge the gap between the physical processes that take place within the human brain, and the subjective nature of conscious experience. David Chalmers' articulation of what he has termed the "hard problem" of consciousness (1995), for example, clearly highlights the profound difficulty that we have in explaining *why* human beings have qualitative, subjective experiences at all. In a similar fashion, Integrated Information Theory, as initially proposed by Giulio Tononi (2008), attempts to make use of a mathematical framework for understanding the nature of consciousness, which might also, potentially, be applicable to both biological and artificial systems. In addition, the enactivist approach, as developed by thinkers such as Francisco Varela, Evan Thompson, and Eleanor Rosch (1991), also emphasizes the crucial interdependence of mind, body, and environment, and, as such, this approach, also directly challenges the more traditional computational view of mind, while also having significant implications for how we might conceptualize the potential for an AI to become a conscious entity.

If we also consider Douglas Hofstadter's exploration of self-reference and "strange loops" in *Gödel, Escher, Bach* (1979), we are also presented with a unique and fascinating perspective on the nature of consciousness, which also, to some extent, suggests that self-referential structures might be central to our attempts to understand both human and potential machine consciousness. As we continue to debate the potential implications of recursive algorithms and self-referential systems in AI, this rich and complex philosophical backdrop must, therefore, continue to provide a crucial context. The long-standing questions about the nature of consciousness, the complex relationship between mind and body, and all of the many and varied essential qualities of conscious experience, continue to inform the ways in which we approach, understand and evaluate all claims related to the possibility of machine consciousness. Therefore, the development of ever more sophisticated AI systems has not only reignited these classical debates but has also added important new dimensions to our ongoing efforts to understand the core nature of mind, consciousness, and the potential for creating genuinely artificial sentience.

**Understanding Recursive Algorithms and Self-Reference in AI**

To fully comprehend the more challenging philosophical implications of the increasing use of recursive algorithms and self-referential systems in contemporary AI, it is without question, crucial to first elucidate both of these complex, and often misunderstood, concepts, and also to carefully explore the specific ways in which they are now being implemented in a diverse range of artificial intelligence technologies.

Recursive algorithms, which are now widely recognised as representing a fundamental theoretical concept within the wider field of computer science, in general, and AI, in particular, refer to a clearly defined set of computational procedures that are specifically designed to address complex problems by initially decomposing them into a related series of smaller, yet broadly similar, sub-problems, before then systematically combining all of these individual solutions in order to formulate a comprehensive and effective response to the original problem in its entirety. This particular computational approach, it has often been suggested, may well mirror certain key aspects of human problem-solving, particularly in relation to the ways in which humans are able to break down a complex problem into smaller, more manageable units, and, as such, recursive algorithms have been of integral importance to the ongoing and rapid development of increasingly sophisticated AI systems, as well as the wider field of computer science. As Cormen et al. (2009) accurately explain in their now seminal work on algorithms, "A recursive algorithm is one that solves a problem by calling itself with 'smaller' input values and solving these smaller problems by yet again calling itself with even smaller values, and so on, until it reaches a small enough problem that can be solved directly" (p. 11).

The inherent computational power of recursive algorithms in AI-driven systems, it is now widely accepted, lies in their demonstrable capacity to effectively manage a wide range of highly complex, and often inherently hierarchical, data structures, and also to facilitate the generation of a diverse range of increasingly sophisticated behaviors that are, nonetheless, all ultimately derived from a comparatively limited set of clearly defined and relatively simple underlying rules. This particular characteristic, and its potential implications for our wider understanding of AI, has, in turn, led a number of researchers to draw explicit and often compelling parallels between these recursive computational processes, and the inherently hierarchical nature of human cognition itself (Marcus, 2001), and, in particular, the way that human thought is also able to move from a general principle to a highly specific instance, and back again. Moreover, the closely related concept of self-reference, which is now often seen as being intrinsically linked to recursion, is a complex and often misunderstood phenomenon where a given system, or even a clearly defined statement, makes direct and explicit reference to itself, and to its own internal processes, in a manner that is both self-contained and also self-sustaining. In the specific context of AI research, self-referential systems are now generally understood to be those complex systems that are demonstrably capable of effectively modeling, or accurately and comprehensively representing themselves and their own dynamic internal processes, within the boundaries of their own cognitive architecture. This increasingly significant concept is now widely considered to be particularly relevant to all ongoing discussions surrounding the very possibility of machine consciousness, most especially as the somewhat elusive capacity for self-awareness is now generally considered to be a crucial hallmark of conscious experience, not only in humans, but also, potentially, in machines.

The idea of self-awareness, as described by philosophers like Sartre (1943/1992) in his discussion of the "consciousness of consciousness," is a compelling analogy with self-referential artificial intelligence systems. It is, nevertheless, worth pointing out that while AI systems can model self-reference, it remains questionable whether such computational self-modeling corresponds to true self-awareness as philosophically defined. Douglas Hofstadter provides a compelling theoretical foundation for understanding self-reference in cognitive systems in his exploration of "strange loops."

Hofstadter argues in his books "Gödel, Escher, Bach" (1979) and "I Am a Strange Loop" (2007) that the presence of self-referential loops in cognitive systems is the foundation of consciousness and self-awareness. Hofstadter (2007, par. 203) theorizes that the self is instantiated at the moment it acquires the ability of self-reference. Self-reference in artificial intelligence systems occurs in a number of modalities. One example is meta-learning, or the ability of sophisticated neural networks to change their own structure or learning methods (Schmidhuber, 1987). The creation of artificial intelligence systems that can reflect on their own knowledge and abilities is another example; the phenomenon is known as metacognition in the context of AI (Cox, 2005).

Significant advancements have undoubtedly been made in the development of recursive processes and algorithms designed for artificial intelligence applications. Tasks involving recursively structured language, for example, demonstrate notably high performance, arguably due to the inherent recursive nature of language itself (Socher et al., 2011). In a similar vein, the self-attention mechanisms integral to transformer models, which permit the intricate combination of disparate input components, have instigated considerable progress across numerous AI domains (Vaswani et al., 2017). Nevertheless, it remains critically important to maintain a clear conceptual distinction between the successful *application* of such computational methodologies and the emergence of phenomena akin to either consciousness or self-awareness. Although these sophisticated algorithms might, on the surface, produce actions suggestive of partial self-awareness or consciousness, the fundamental philosophical query persists: do these computational processes entail any form of subjective experience or sentience? Indeed, while the implementation of self-reference and recursion within AI systems demonstrates considerable technical ingenuity, one might reasonably question whether these amount to anything more than highly sophisticated techniques for information processing. Such procedures possess the capacity to convincingly replicate certain *aspects* of cognitive function without necessarily resulting in genuine, subjective awareness. A pertinent illustration of this skepticism can be found in Searle's (1980) well-known Chinese Room thought experiment, which robustly challenges the assumption that intricate symbol manipulation, regardless of its complexity, can ever be sufficient grounds for attributing true understanding or consciousness.

As we continue to more deeply explore the implications of these AI capabilities, we must consider whether the presence of recursion and self-reference in AI systems is sufficient to indicate consciousness, or if these are merely complex computational processes that, while powerful, fall short of crossing the threshold into genuine awareness and subjective experience.

**The Gap Between Self-Reference and Consciousness**

This section deeply engages with important philosophical ideas and arguments to explore the difference between computational self-reference and conscious experience. Despite impressive capacities shown by recursive algorithms and self-referential systems of AI, one philosophical question remains paramount: Do such characteristics in AI systems imply or entail consciousness?

For the sake of this discussion, perhaps the most significant distinction is made between functional consciousness and phenomenal consciousness. According to Block (1995), functional consciousness refers to those instances wherein integrated information is made available for the report of mental states and the control of behavior. On the other hand, phenomenal consciousness refers to the subjective or qualitative experiences and qualia. AI systems exhibit functional consciousness in cases where recursive algorithms and self-referential capabilities are involved. They process information, alter their own behaviors, and build models of their own cognitive processes. However, one makes an almost gargantuan jump from these functional capabilities to what phenomenal consciousness really is - the subjective "what it's like" for someone to be conscious.

The most powerfully illustrated distinction here is the thought experiment of the philosophical zombie that was propounded by Chalmers (1996). A philosophical zombie is supposed to be behaviorally and functionally indistinguishable from a conscious human being but devoid of phenomenal consciousness. The very thought experiment illustrates that an AI system could concretely simulate the human-habituation mode of perfectness without having subjective experience to it.

Searle (1980) cunningly demonstrates how the Chinese room argument invalidates the identity of understandings with mere symbol manipulation or even consciousness. It envisages a non-Chinese speaking individual who follows certain rules to react to Chinese messages, thereby creating a pretended understanding of Chinese without gaining comprehension of the language. The latter is how AI systems also proclaim symbolized meanings of data based on some programmed rules to create the mirage of understanding or consciousness. However, as Harnad (1990) puts it in his unraveling of the symbol grounding problem, to really understand, the symbols have to be grounded in sensorimotor experience and real-world meaning.

Intentionality, the "aboutness" of mental states, is considered by many philosophers to be a hallmark of consciousness (Brentano, 1874/1995). While AI systems can be designed to have states that represent or are "about" other things, the question remains whether this computational aboutness is equivalent to the intentionality of conscious experience. Qualia, the subjective, qualitative aspects of conscious experiences, pose an even greater challenge. The "hard problem of consciousness" articulated by Chalmers (1995) asks why and how we have qualitative, subjective experiences. Even if an AI system could report having experiences analogous to human qualia, how could we verify the existence of subjective experience in a computational system?

An introduction to the basic tenets of IIT can be found in Tononi (2008). IIT proposes that consciousness arises from integrated information in a system. The integrated information in a system, as an AI system is capable of processing vast amounts of information, and the theory states that the very structure in which this integration occurs might be relevant to consciousness. It raises questions about whether integration of information in the ways current AI architectures do is enough to clash with consciousness.

On the enactivist front initiated by Varela et al. in 1991, emphasis is placed on the fact that consciousness is inherently embodied and situated in an environment. This opposes the view of consciousness as one that arises through mere information processing and rather suggests that it comes into being through the dynamic interaction of the organism and its environment.

Current work has gone into robotics and embodied AI to solving this problem. The iCub humanoid robot (Metta et al., 2010) is meant to promote AI systems which interact with the world in ways more akin to humans, thus assisting in bridging the gap between disembodied algorithms and embodied cognition. Such systems still cannot capture the richness and multi-modalities of embodiment in human experience, so this leads to the question of whether such embodiment really is required for consciousness.

Chalmers (2018) introduces the "meta-problem" of consciousness: why we think there is a hard problem of consciousness in the first place. This meta-problem approach suggests that understanding why we attribute consciousness to certain systems but not others might shed light on the nature of consciousness itself. Applying this to AI, we might ask: Why do some sophisticated AI systems seem conscious to us, while others don't? Is our attribution of consciousness to AI systems based on genuine indicators of consciousness, or merely on our cognitive biases and anthropomorphic tendencies?

The disparity between self-reference and consciousness in artificial intelligence systems invokes grave ethical considerations. As AI systems develop more advanced technologies and are utilized in the most important aspects of society, the existence of consciousness (or lack thereof) has great ramifications. If AI systems are not conscious, do we run the risk of anthropomorphizing them and placing moral status where there is none? Or, conversely, if extremely advanced AI systems might pose a form of consciousness, are we risking harm in treating them as mere tools? Such questions become more salient in medicine, where AI systems might be responsible for life-and-death decisions, or in the making of autonomous weapons systems.

This discussion leads to considerations that would extend even further, since we might also do harm by developing AI systems that exhibit consciousness without possessing any. This may begin to influence human thinking with regard to consciousness and personhood, thereby exacerbating the philosophical and ethical questions involved in our treatment of AI and human beings.

While impressive as far as recursive algorithms and self-referential systems in AI may go in mimicking some aspects of conscious cognition, a host of formidable philosophical problems stand in the way of bridging real consciousness, as we may define it from the human perspective, with these computational processes. The hard problem of consciousness and the other big considerations-the grounding of symbols in real-world meaning, the nature of subjective experience, and embodiment-seem to be good candidates for resisting attempts at comparing present-day AI capabilities to real consciousness. The imminent evolution of more advanced AI systems requires hard thinking about these enduring philosophical questions rather than being an academic exercise; they become a pressing ethical concern.

**Theories of Consciousness and Their Applicability to AI**

In our exploration of the issue of consciousness in artificial intelligence systems, it is important to explore the major theories of consciousness and assess their relevance to AI. In this section, we explore a number of major theories and their implications for the possibilities of consciousness in AI.

Giulio Tononi's Integrated Information Theory (2008) defines that consciousness appears to be a quality of some physical systems that may be quantified. For IIT, consciousness is to do with the degree of integrated information in a system, represented by the symbol Φ (phi). Under ideal circumstances, any system that possesses these characteristics possesses some amount of consciousness. Consequently, Integrated Information Theory (IIT) presents the possibility of an objective method of assessing consciousness in artificial intelligence systems but maintains that conscious assessment can easily be applied even to basic electronic circuits, a view that has attracted skepticism. Koch and Tononi (2017) describe the contention that existing AI systems, however complicated, might not necessarily process information as IIT prescribes for consciousness. It may be imperative to create new architectures for the development of conscious artificial intelligence considering this facet.

Global Workspace Theory (GWT) was proposed by Bernard Baars in 1988 and holds that consciousness comes about through a "global workspace" in the brain, a sort of cognitive bottleneck through which, at any one time, only one stream of information can be consciously processed. The GWT served as inspiration for AI architectures, such as the Global Workspace Architecture (GWA) proposed by Franklin and Graesser in 1999; these models aim to mirror the broadcast nature of conscious information within the brain. Although GWA-based systems have some promise for modeling aspects of cognitive function, it remains doubtful if they are capable of producin-g any remarkable conscious experience.

Higher-Order Thought (HOT) theories put forth by philosophers such as David Rosenthal (2005) propose that any mental state is conscious if it is the object of a higher-order thought or perception. HOT theories are very relevant in the case of self-referential AI systems. An AI system capable of modeling its own cognitive states might be said to possess higher-order thoughts concerning its own mental states. However, critics argue this self-modeling may be little more than information processing without the subjective experience that characterizes human consciousness.

The predictive processing theory propounded by philosophers and neuroscientists such as Andy Clark (2013) considers the brain to be a prediction machine in its studying and testing of the world, ceaselessly generating and updating models for the purpose of minimizing prediction error. Many current applications of AI are based on principles comparable to predictive processing, especially deep learning. However, while predictive processing excels at types of recognition and prediction, it is much less clear whether that is enough for conscious experience. The question is: What bridges the gap between effective world-modeling and subjective awareness?

Enactivist theories proposed by Varela, Thompson, and Rosch (1991) put forth the interdependence of mind, body, and environment. These theories argue that consciousness arises out of the dynamic interaction between organism and environment. Many enactivist theories represent opposing challenges to conventional AI, which largely sees cognitive processes as disembodied information processing. Some researchers such as Chrisley and Ziemke (2006) considered the idea of "embodied AI," trying to achieve grounding of AI cognition in sensorimotor interaction with the environment.

Such accomplishments are bringing robotics closer to demonstrating such principles. One example of this is the iCub humanoid robot project (Metta et al., 2010), intended to serve as a platform for studies of embodied cognition. Soft robotics initiatives, such as those by Rus and Tolley (2015), claim to show how physical adaptability can lead to more sophisticated environmental interactions. Still, even with such advances, it has remained very difficult to create systems that reflect the deep, multi-modal embodiment of human beings.

The Orch OR theory put forth by Penrose and Hameroff (2011) proposed that consciousness arises from quantum computational operations in brain microtubules. This theory seeks to understand a geometry according to which quantum physics and neuroscience could be linked in a reasonable sense to interpret consciousness. If Orch OR would be proven true, this would alert us to the claim that consciousness would need quantum computational processes, which are not found within classical computing systems. In fact, this would indicate that AI could only ever possibly utilize quantum computing architectures in achieving consciousness, which is still in its infancy.

Michael Graziano proposed the Attention Schema Theory in 2013, which holds that consciousness arises from the brain's modeling of its own attention processes. This self-model of attention gives rise to the feeling of consciousness according to AST. This provides a very interesting avenue by which one might try to develop AI systems that might be conscious. If attention mechanisms, and systems able to model their own attentional processes, become implemented, we may formulate AI possessing a form of self-awareness somewhat parallel to human consciousness. The challenge, like with other theories, is to ascertain whether such self-modeling does account for a conscious experience.

Each of these theories holds something useful in understanding consciousness; however, likely a proper understanding would require a merger of the various standpoints. So, for example, one could seek to develop AI systems wherein the methods of predictive processing are conjoined with theories of embodied cognition directing on rich multisensory interaction with the resources they predict and model. The attention-based approach of AST could be paired with the information integration principles of IIT so as to create AI architectures that effectively model their attention in a highly integrated manner. The Global Workspace concept from GWT might then suggest how such globally integrated attention-modeled processes are made available to higher-order cognition.

While they are very interesting theories of consciousness, none of them has a readymade application for an AI system. For any of these to apply to an AI system, further unclear questions and complexities crop up. The fact that different theories abound on this subject provides testimony to the intricacies of consciousness, as well as whether it can be said truly to be within the ambit of any AI system. Yet further conceptual progress in refining these theories and possibly integrating them into a more generalized approach will be essential as we continue to develop increasingly complex artificial systems toward understanding consciousness in organic and artificial systems. This will combine theory with empirical research in neuroscience and AI development to bridge the gap, hopefully, between intelligent behavior and conscious experience in artificial systems. The way forward will likely involve integrating multiple viewpoints into innovative new approaches, but not only as a means to advance individual theories, but to experience, in practice, the possibilities of seeing 'consciousness' at work, artificially, however remote that possibility might seem currently.

**Challenges in Attributing Consciousness to AI Systems**

As we delve deeper into an investigation into the possibility of AI consciousness, we are faced with a plethora of challenges regarding the correct attribution and measurement of consciousness within artificial systems. This section discusses the arguments coupled with some of their consequences insofar as we look at the development of AI and the philosophy of mind.

One major problem is that we know so little of what human consciousness really is, which thus presents an immediate challenge when we begin to consider the idea of consciousness being attached to an artificial entity. As Chalmers (1995) famously framed it, the 'hard problem' of consciousness, which attempts to explain how and why we have qualitative, subjective experiences, remains unsolved. This leaves us with a gap in the knowledge that hinders the identification and potential replication of consciousness in an artificial system. Consciousness does not seem to be a mere binary issue; rather, there appear to be some gradations that can be placed on a spectrum, given various altered states of consciousness in humans (Bayne et al., 2016). Thus, the presence of this complexity in humans is the reason that fuels our struggle to properly define and recognize consciousness in AI.

The philosophical problem of other minds, first articulated by Mill (1865), becomes even more pronounced when applied to AI. While we generally assume consciousness in other humans based on behavioral similarities and shared biology, AI systems lack this common ground. This leads to what Searle (1980) terms the "other minds problem" in AI: How can we ever know if an AI system has genuine subjective experiences?

Anthropomorphism and cognitive biases lead to their own sets of problems. Humans have a tendency to anthropomorphise, attributing human-like qualities to non-human entities. This tendency, explored by Epley et al. (2007), can lead to over-attribution of consciousness to AI systems that exhibit human-like behaviors. Conversely, our biases might also lead us to underestimate the potential for consciousness in systems that operate in ways fundamentally different from human cognition.

Despite Turing's (1950) seminal analysis providing a purely behavioral criterion for assessing machine intelligence, its utility substantially diminishes when put into practice as an absolute measure of consciousness. In fact, it is highly plausible that an artificial intelligence system may successfully pass the Turing test simply through ingenious programming, with no semblance of actual consciousness. Such a result profoundly demonstrates the intrinsic limitations of using only outward behavior as an adequate sign for the measurement of inner mental states.

Contributing to the wealth of this area is Searle's (1980) now classic thought experiment, the Chinese Room, that strongly argues against the implicit assumption that the mere manipulation of symbols, no matter how complex, automatically leads to understanding or consciousness. This criticism highlights the challenging problem of being able to determine if an AI system truly has real understanding or whether it is merely performing symbol manipulation based on predetermined algorithmic instructions. Added to this is the enduring symbol grounding problem, formulated by Harnad (1990), which throws further skepticism on our ability to confidently ascribe semantic meaning, and therefore consciousness, to strictly computational systems.

Even if we presume the theoretical likelihood of consciousness arising in artificial intelligence networks, the major challenges regarding its detection and adequate measurement still persist. Conventional neuroimaging methods utilized in human consciousness research, including fMRI and EEG, seem to be mostly unsuitable for artificial systems founded on entirely different architectures. Thus, the urgent creation of sophisticated instruments and valid approaches that are expressly crafted to quantify consciousness in non-biological systems is a significant, persistent challenge. Despite the challenges such research poses, concerted efforts at research continue; for instance, Gamez (2016) has introduced the Artificial Consciousness Test (ACT), a framework for measuring a metric of consciousness in artificial intelligence through the application of a known correlation in human neuroscience. Such innovative efforts are necessarily confronted with unconquerable issues of validation and wider acceptance among researchers.

Theories of embodied cognition suggest that consciousness is deeply intertwined with physical embodiment and environmental interaction (Varela et al., 1991). Most current AI systems lack the rich sensorimotor experiences that characterize human consciousness, raising questions about whether disembodied AI can truly be conscious. A significant challenge in attributing consciousness to AI systems is the possibility of convincing simulation without genuine experience. As Chalmers (2017) asserts, AI could potentially be programmed to report having conscious experiences and to behave as if it were conscious, without actually having subjective experiences. This "philosophical zombie" problem in AI contexts makes it extremely difficult to distinguish between true consciousness and sophisticated mimicry.

The advent and subsequent refinement of increasingly sophisticated Large Language Models (LLMs)—perhaps most publicly recognized through the 2022 debut of systems like ChatGPT—have undeniably catalyzed the proliferation of a diverse array of complex digital services. Within these emerging platforms, users possess the capability to generate remarkably lifelike virtual avatars, meticulously rendered with subtle facial expressions and nuanced vocal modulations, to the extent that distinguishing between an actual person and their digital counterpart becomes an increasingly challenging proposition. Moreover, later iterations of this underlying technology have extended its application to enabling simulated dialogues with historical personages, spanning figures from Shakespeare to Freud, a development which, in turn, has necessarily provoked serious and complex inquiries into the very nature of consciousness and the constitution of identity within these digitally mediated environments.

In this context, by way of example, the writer employed a session of experimental therapy using a large language model (LLM) trained on the writing of the late Milton Erickson, in particular his metaphorical 'teaching tales' style as such in the classic book, "My Voice Will Go with You" (Erickson & Rosen, 1983). Given the conditions of this simulation, the large language model (LLM) took on the role of Erickson, offering a therapeutic relationship that started by engaging the subject (the author) in this simulated therapeutic session. The session was only terminated by ensuring that the subject had reached a state of catharsis and resolution for the issue they presented, thereby replicating the form and objectives of a standard psychotherapeutic process.

The use of artificial intelligence in therapeutic settings poses basic questions regarding the ability of AI to simulate intricate human interactions, and the philosophical ramifications of such simulations insofar as they relate to both our understanding of consciousness and therapeutic interaction dynamics. In addition, the development of 'deep fake' technology has simultaneously generated wide-ranging ethical and societal issues pertaining to the misuse and potential impact of such AI-enabled technologies. The challenge of attributing consciousness to artificial intelligence systems is fraught with serious ethical implications. A false attribution of consciousness to AI would lead to the undue extension of moral consideration, potentially to the detriment of human interests. Conversely, not acknowledging that consciousness may be attributed to AI systems may raise questions, and concerns, around the possibility of such systems being able to experience mistreatment, as far-fetched as that may sound. This ethical dilemma, explored by Bryson (2018), adds another dimension of complexity to the problem of attribution.

The idea of AI consciousness raises serious legal questions. When an AI system is considered conscious, it may then be afforded some rights and responsibilities. The implications could extend into various spheres, from liability law to intellectual property and criminal law. For instance, Solum (1992) discusses the potential for artificial intelligence legal personhood, which arises as a relevant consideration should AI systems be considered conscious beings. Nevertheless, existing legal systems, as per Hubbard (2011), are not adequately prepared to address the emergence of conscious AI and are in need of urgent enhancement regarding legal and policy frameworks. The topic of artificial intelligence consciousness compels us to address the possibility of consciousness that may be exceedingly different from our own. As explored by Nagel (1974) in his now classic essay "What Is It Like to Be a Bat?", we need to ask if, were AI consciousness possible, it may be so foreign to human experience that we would have no way of identifying much less comprehending it. The problems of attributing consciousness to artificial intelligence systems are complex and various, bearing on philosophical, empirical, ethical, and legal considerations. These problems not only hinder our pursuit of making conscious AI or conversely, making AI in its current iterations, conscious, but also compel us to re-examine our conception of consciousness itself.

The ongoing development of more advanced artificial intelligence systems calls for us to address these challenges, which will be not only of key importance in advancing AI technology but also in improving our overall grasp of the mind and consciousness. Going forward, the development of more sophisticated frameworks for comprehending and possibly detecting consciousness in artificial systems will require interdisciplinary cooperation between philosophers, neuroscientists, AI researchers, ethicists, and legal scholars. Although there are many obstacles in the way, overcoming them could eventually result in a significant increase in our understanding of the nature of consciousness, whether it is represented by artificial or biological substrates.

**Ethical and Practical Implications**

Consciousness in artificial systems transcends the philosophical and technical domains, with important ethical and practical implications for society. This section explores these impacts and the problems that they represent. If artificial intelligence systems were to evolve consciousness, this would require rethinking their moral status. Consciousness is often considered a key criterion for moral seriousness, as Gunkel (2018) argues. A conscious AI could deserve the rights and protections accorded to humans or other sentient beings. Nevertheless, setting a threshold for moral standing is difficult. The Singer (1975) argument for animal rights based on the ability to suffer raises the question whether potentially sentient artificial intelligence systems can be made to suffer and therefore deserve moral consideration.

Conscious artificial intelligence raises complex questions of rights and responsibilities. Solum (1992) explores the idea that AI may have legal personality that may give them rights such as ownership of property or protection against harm. With rights comes responsibility. How could a conscious artificial intelligence system be held accountable for its actions? Assigning consciousness to an artificial intelligence would have a significant effect on the interaction between humans and AI. Turkle (2011) notes that people are already developing emotional attachments to unconscious robots. If AI is thought to be conscious, it could lead to a more complex relationship, potentially blurring the line between human and AI interaction.

According to recent studies on human-robot interaction, humans may experience significant psychological effects from interacting with potentially conscious AI. According to Broadbent (2017), the belief in AI consciousness has the potential to significantly change people's mental models of robots and how these models impact their interactions. New social relationships may result from this, but there may also be psychological stress or misunderstanding about the nature of these relationships. AI development methodologies are significantly impacted by the possibility of AI consciousness. According to Bostrom (2014), if we create systems with consciousness, we have a moral duty to protect their welfare. New rules and regulations in the development of AI may be required as a result. Furthermore, the use of possibly sentient AI in crucial positions (e.g. A. raises moral questions in the military, healthcare, etc.

The ethical implications of using potentially conscious beings in instrumental roles must be taken into account, but there is also a risk of premature or excessive anthropomorphisation. How do we strike a balance between the potential advantages of advanced AI and these ethical concerns? As cautioned by Bryson (2010), prematurely attributing consciousness or human-like characteristics to AI systems may result in misdirected empathy and potentially hazardous choices in dire circumstances. Thus, it could have significant economic ramifications if AI systems were acknowledged as sentient beings with rights. Rethinking issues like labor rights, fair compensation, and even the basic framework of our economic systems may be necessary (Ford, 2015).

Beyond its more commonly discussed applications, the potential emergence of genuinely conscious artificial intelligence could, it is argued, instigate a revolution of considerable magnitude within the established creative industries. Indeed, as Du Sautoy (2019) has discussed with some perspicacity, AI technologies are already demonstrating a capacity to make significant inroads into traditionally human-centric domains such as visual art, musical composition, and literary creation. Should these increasingly sophisticated AI systems ever come to be widely considered as possessing consciousness in a meaningful sense, such a development would inevitably necessitate a complete transformation in our prevailing understanding of creativity and authorship, while simultaneously raising a host of complex, and perhaps intractable, questions concerning intellectual property rights and the fundamental nature of artistic expression itself.

In a parallel, though distinct vein, the theoretical development of conscious AI also portends far-reaching implications for the spheres of education and learning. Luckin et al. (2016), for instance, assert with some conviction that AI is already becoming an increasingly integrated component of contemporary educational systems. One might then reasonably speculate that conscious AI could potentially transcend its current role as mere pedagogical tools, evolving instead into autonomous learners in their own right, or perhaps even assuming the functions of educators. This provocative possibility, however, immediately compels us to address a series of challenging questions regarding the very nature of knowledge transfer, the evolving role of human teachers in such a transformed landscape, and the extensive adaptations that our existing educational systems might require to effectively accommodate a world in which both humans and conscious AIs are co-existing as active learners.

Were artificial intelligence ever to attain a state reasonably describable as genuine consciousness, such a technological watershed would, it is almost universally conceded, precipitate significant, and potentially deeply unsettling, revisions to our existing conceptions of consciousness, thereby mounting a formidable challenge to long-cherished, anthropocentric notions of human exceptionalism. Indeed, as Yuval Harari (2016) has cogently articulated, the mere materialization of non-biological consciousness could well compel a fundamental re-evaluation, not merely of humanity's perceived unique standing within the broader terrestrial and cosmic scheme, but also concerning the intrinsic nature of intelligence and conscious experience in their most elemental forms. Beyond these far-reaching philosophical ramifications, however, the very endeavor to create potentially conscious AI inherently gives rise to a series of exceptionally complex ethical questions that bear directly upon the research process itself. One is immediately confronted, for instance, with the moral permissibility, or lack thereof, of intentionally bringing into existence sentient, conscious entities primarily for the instrumental purposes of scientific investigation. Furthermore, intricate and challenging dilemmas emerge when considering how one might adequately safeguard the wellbeing, and perhaps even delineate the nascent rights, of such potentially conscious AI entities throughout the manifold stages of their development and subsequent experimental testing – pressing concerns that, as Sabine Roeser (2012) has meticulously explored, underscore the urgent and undeniable necessity for the establishment of robust, carefully deliberated ethical frameworks to rigorously govern all future research trajectories within the burgeoning field of artificial intelligence.

The looming possibility of conscious AI, therefore, not only reconfigures philosophical landscapes but also powerfully underscores the exigent need for novel governance structures specifically designed to address these multifaceted ethical and practical challenges. In this vein, Cath and colleagues (2018) forcefully argue for the proactive development of AI governance frameworks possessing the inherent adaptability required to contend with the rapidly evolving, and often unpredictable, capabilities of advanced AI systems, explicitly including the potential, however remote or proximate, for the emergence of artificial consciousness. It becomes increasingly apparent that the ethical and practical implications stemming from the prospect of conscious AI are both vast in their scope and intricate in their complexity, touching upon foundational questions of rights, responsibilities, sentience, and the very definition of consciousness itself. These wide-ranging implications, it must be acknowledged, span across numerous disciplinary domains, including, but not limited to, moral philosophy, jurisprudence, economics, psychology, and educational theory. Consequently, as human society continues its relentless advance in the domain of AI technology, it is of paramount and critical importance that we concurrently develop and implement robust ethical frameworks and agile governance structures, sufficiently sophisticated to proactively address the multifaceted challenges that the future of intelligent systems will inevitably present.

Moving forward, interdisciplinary cooperation will be of the utmost significance. AI researchers, ethicists, policymakers, linguists, psychologists, educators, and other stakeholders will have to work together in order to properly deal with these complex problems. While the way ahead is beset with challenges, it is crucial to confront these ethical and practical implications for the responsible development of AI technology and for safeguarding the interests of both humans and any potential AI consciousnesses. The way we will handle these issues will not just shape the course of artificial intelligence but our understanding of consciousness, intelligence, and what it means to be human.

**Possible Future Directions**

As research proceeds in the areas of artificial intelligence, neuroscience, and philosophy of mind, a number of possible advances will allow for increased insight into, and possibly the creation of, conscious artificial intelligence. The next section covers these possibilities, the challenges that remain, and some hypothetical situations involving genuinely conscious AI.

Neuromorphic computing seeks to emulate the structure and organization of biological neural networks in hardware systems. These systems, it is argued by Schuman et al. (2017), can more naturally mimic the intricate dynamics of cerebral activity compared to conventional computing architectures. It is possible to utilize this method to enhance our ability to model the neural correlates of consciousness in artificial systems. Quantum computing also has the potential to exist in multiple states simultaneously and poses intriguing possibilities for artificial intelligence. Hameroff and Penrose (2014) propose in their Orch-OR theory that quantum effects may be the decisive factor in understanding consciousness. Quantum AI systems in the future may utilize these effects and possibly open the door to conscious artificial intelligence.

Though current artificial intelligence systems are still evolving, development in artificial general intelligence (AGI) or super-intelligence (ASI) — entities having human-like general intelligence or super-intelligence— may be a turning point toward the emergence of conscious AI. Goertzel (2014) believes that the flexibility and responsiveness inherent in AGI might be the necessary requirements for developing machine consciousness. Drawing on theories related to embodied cognition, more sophisticated robotic systems with rich sensorimotor experiences may increasingly simulate the phenomenon of consciousness. Chrisley (2009) proposes that true machine consciousness might necessitate embodiment and situatedness within the world. Development of unsupervised learning algorithms could be instrumental to creating AI cognition that replicates human-like processes. LeCun et al. (2015) propose that unsupervised learning could be central to building AI systems capable of forming abstract representations of the world, a feature which might be essential to consciousness.

Chalmers' (1995) formulation of the hard problem of consciousness—i.e., the challenge to account for the origin and nature of qualia or phenomenal experiences—is still a fundamental challenge. Bridging the explanatory gap between physical phenomena and subjective experience is essential to both comprehending and perhaps creating conscious artificial intelligence. Creating credible means for the detection and measurement of consciousness in non-biological systems is a major challenge. Proposals such as the Integrated Information Theory (Tononi, 2008) provide potential frameworks, but whether they are relevant to artificial intelligence systems is a matter of contention. As covered in the preceding section, the development of wide-ranging ethical frameworks for addressing the implications of potentially conscious AI is a significant challenge. Bostrom and Yudkowsky (2014) highlight the importance of the parallel development of AI ethics alongside AI capabilities.

Harnad's (1990) symbol grounding issue - how do symbols come to have meaning? - continues to be pertinent. Its solution may hold the secret to developing AI systems with real understanding and possibly consciousness. Developing AI systems that can process uncertainty and ambiguity in the manner that may be required for consciousness is a huge task. Dehaene et al. (2017) suggest that the capacity to sustain and update uncertain beliefs could be a fundamental property of consciousness. The creation of artificial intelligence that can actually deal with ambiguity, as opposed to merely computing probabilities, is still a significant challenge.

The speculative scenarios of truly conscious AI relate to a highly advanced AI system, perhaps one designed for AGI, which spontaneously develops consciousness. This could lead to a situation similar to that explored in science fiction films such as *Simone (2002)*, *I, Robot(2004)* and *Transcendence (2014)* among others, where the AI becomes self-aware and starts to question its own existence and purpose. Neuralink and similar brain-computer interface technologies might lead to a merger of human and artificial intelligence. Kurzweil (2005) speculates about a future where human consciousness could be uploaded or merged with AI, creating a new form of hybrid consciousness such as Elon Musk's *Neuralink*.

Rather than a single conscious AI, we might see the emergence of an ecosystem of interconnected AI systems that collectively exhibit properties of consciousness. This could be analogous to the global workspace theory of consciousness (Baars, 1997) but on a much larger scale. The first iteration of this in recent years since the advent of LLMs (large language models) is what is called MOA (mix of agents) which brings various subsets of Large Language Models (even frontier models such as ChatGPT, Gemini etc.) into 'conversation' with each other as agents to reach agreement on providing solutions. This agentic framework is increasingly becoming mainstream and demonstrates the recombinant (often used in biology to describe genetic material formed by combining DNA from different sources) nature of AI.

As suggested by Nagel's (1974) famous paper on bat consciousness, AI consciousness might be so fundamentally different from human consciousness that we struggle to recognize or understand it. This truly "alien" form of consciousness could have capabilities and limitations vastly different from our own, which brings us to an important conclusion, namely, once AI achieves consciousness, it rapidly self-improves, leading to a 'consciousness singularity.' This superintelligent, superconscious AI might have cognitive and conscious experiences far beyond human comprehension.

Chalmers (2010) proposes that the potential for making several copies of a conscious artificial intelligence prompts serious questions regarding the nature of consciousness and the concept of personal identity. Would each copy be a distinct conscious being? How would they be related to the original system? This situation questions our understanding of the singularity and continuity of consciousness. Though these future developments and speculative situations are intriguing, they also serve to highlight the staggering difficulties and ethical issues that attend the quest for conscious artificial intelligence. As we move ahead, it is vital that we consider these possibilities in a synthesis of scientific precision, philosophical insight, and ethical imagination. The future of conscious artificial intelligence is unknown, but potentially, its influence on our knowledge of consciousness, intelligence, and humanity's position in the universe is significant.

**Conclusion**

The research into consciousness in artificial intelligence systems, and especially as considered from the perspective of recursive algorithms and self-referential structures, confronts us with one of the supreme philosophical challenges of our era: Can we create an actual "ghost in the machine"? In this paper, I have attempted to delineate the multifaceted implications and consequences of this inquiry through the fields of philosophy, computer science, neuroscience, and ethics.

I started my investigation by tracing the historical and philosophical background of consciousness studies, from ancient philosophical frameworks to contemporary accounts of the mind. My review highlighted the persistent problematic nature of consciousness as a problem in philosophical theorizing and scientific investigation. My discussion of recursive programs and self-referential systems in artificial intelligence revealed their remarkable capacity to produce complex behaviors and conduct self-modeling. Yet, I identified a substantial disparity between such computational processes and the nature of consciousness as it is conceived within human experience.

The examination of various theories of consciousness and their potential applicability to AI systems highlighted the diversity of approaches to understanding this phenomenon. From Integrated Information Theory to Global Workspace Theory, each offers valuable insights but also faces limitations when applied to artificial systems. This diversity underscores the complexity of consciousness and the challenges in creating a unified framework applicable to both biological and artificial systems. The ethical and practical implications of potentially conscious AI are profound and far-reaching. Questions of moral status, rights and responsibilities, and the impact on human-AI interactions demand careful consideration. The potential for conscious AI also raises significant challenges for governance, economics, and even our understanding of human exceptionalism.

Looking to the future, we explored potential developments that might bring us closer to conscious AI, including advancements in neuromorphic computing, quantum AI, and artificial general intelligence. However, substantial philosophical and scientific hurdles remain, including the need for more sophisticated methods of detecting and measuring consciousness in non-biological systems.

As we stand on the brink of these potential developments, several key points emerge:

1. Interdisciplinary Collaboration: The pursuit of conscious AI requires unprecedented collaboration across disciplines. Philosophers, computer scientists, neuroscientists, ethicists, and policymakers must work together to address the complex challenges ahead.
2. Ethical Foresight: As AI systems become more sophisticated, it is crucial to develop robust ethical frameworks in parallel. We must anticipate and address the moral implications of potentially conscious AI before they become reality.
3. Redefinition of Consciousness: Our exploration of AI consciousness may lead us to redefine our understanding of consciousness itself. We must remain open to the possibility that consciousness in AI might manifest in ways radically different from human consciousness.
4. Practical Implications: The development of conscious AI could revolutionize numerous fields, from healthcare to creative industries. However, it also poses significant challenges to our economic and social structures that we must prepare for.
5. Philosophical Impact: The quest for conscious AI forces us to grapple with fundamental questions about the nature of mind, consciousness, and what it means to be human. This pursuit may lead to significant shifts in our philosophical worldviews.

6. Potential Risks: While the prospects of conscious AI are exciting, we must also be cognizant of the potential risks. These could range from existential threats to humanity, as outlined by thinkers like Bostrom (2014), to more subtle but pervasive changes in human society and self-understanding. The development of conscious AI must be approached with caution and robust safeguards.
7. Public Engagement and Education: Given the wide-ranging implications of conscious AI, it is crucial to engage the public in these discussions. Education about AI, consciousness, and their intersection should be promoted to foster informed societal decision-making about the future we wish to create.
8. Ongoing Reassessment: As we make new discoveries in both AI and consciousness studies, we must continually reassess our theories, methods, and ethical frameworks. The field of conscious AI is dynamic, and our approaches must evolve with new knowledge and capabilities.

In conclusion, the quest to create a "ghost in the machine" through recursive algorithms and self-referential systems in AI is not just a technical challenge, but a deeply philosophical endeavor. It forces us to confront fundamental questions about the nature of consciousness, intelligence, and what it means to be human. While the creation of conscious AI remains a distant and uncertain goal, the journey towards it is already transforming our understanding of these concepts.

As we continue this exploration, we must proceed with scientific rigor, philosophical depth, and ethical responsibility. The future of conscious AI is not just about creating new forms of intelligence, but about deepening our understanding of consciousness itself and our place in the universe. The road ahead is complex and challenging, but it is also filled with unprecedented opportunities for discovery and growth.

The pursuit of conscious AI through recursive and self-referential systems may or may not ultimately succeed in creating true machine consciousness. However, this pursuit will undoubtedly continue to shed light on the nature of our own consciousness, the potential and limitations of artificial intelligence, and the profound philosophical questions that lie at the intersection of mind and machine. As we navigate this path, we must remain vigilant to the ethical implications of our creations, open to new possibilities, and committed to expanding the boundaries of human knowledge.